# Design of p-Type Cladding Layers for Tunnel-Injected UV-A Light Emitting Diodes

Yuewei Zhang,[1,a)] Sriram Krishnamoorthy,[1] Fatih Akyol,[1] Andrew A. Allerman,[2] Michael W. Moseley,[2] Andrew M. Armstrong,[2] and Siddharth Rajan[1,3,a)]

[1] Department of Electrical and Computer Engineering, The Ohio State University, Columbus, Ohio, 43210, USA

[2] Sandia National Laboratories, Albuquerque, New Mexico 87185, USA

[3] Department of Materials Science and Engineering, The Ohio State University, Columbus, Ohio, 43210, USA

**Abstract:** We discuss the engineering of p-AlGaN cladding layers for achieving efficient tunnel-injected III-Nitride ultraviolet light emitting diodes (UV LEDs) in the UV-A spectral range. We show that capacitance-voltage measurements can be used to estimate the compensation and doping in p-AlGaN layers located between the multi-quantum well region and the tunnel junction layer. By increasing the p-type doping concentration to overcome the background compensation, on-wafer external quantum efficiency and wall-plug efficiency of 3.37% and 1.62% were achieved for tunnel-injected UV LEDs emitting at 325 nm. We also show that interband tunneling hole injection can be used to realize UV LEDs without any acceptor doping. The work discussed here provides new understanding of hole doping and transport in AlGaN-based UV LEDs, and demonstrates the excellent performance of tunnel-injected LEDs for the UV-A wavelength range.

---

a) Authors to whom correspondence should be addressed.
    Electronic mail: zhang.3789@osu.edu, rajan@ece.osu.edu

III-Nitride ultraviolet light emitting diodes (UV LEDs) have attracted great research interest for applications including sterilization and water purification.[1] UV light emissions down to 210 nm have been achieved for AlGaN-based UV LEDs.[1,2] However, the wall-plug efficiency (WPE) of UV LEDs remains much lower than InGaN-based visible LEDs, especially in the wavelength range below 365 nm, even though high internal quantum efficiency up to 80% has been demonstrated due to optimized substrate and active region quality.[2,3] The reduced efficiency is attributed to a combination of the poor light extraction efficiency due to high internal light absorption in the widely used p-GaN cap layer and the low injection efficiency associated with the low thermally activated hole density.[4-6] Both originate from the high acceptor activation energy in AlGaN, which leads to poor p-type conductivity and resistive p-type contacts.[5,7]

To solve both the absorption and electrical loss issues, we adopt non-equilibrium interband tunneling for efficient hole injection.[8-15] As a substitute for direct p-type metal contact, a tunneling contact is formed by connecting a transparent n-AlGaN top contact layer to the p-AlGaN layer using a polarization-engineered tunnel junction layer.[10] This leads to minimal internal light absorption due to the absence of the thick absorbing p-type contact layers, making it possible to achieve greatly enhanced light extraction efficiency.[5,6,16] Another advantage is that high tunneling hole current can be achieved at small reverse bias for well-designed tunnel junctions, contributing to enhanced injection efficiency and great reduction in the operation voltage. Since effective hole injection is realized through interband tunneling, which is no longer limited by thermal activation of acceptors, this approach is especially useful for ultra-wide bandgap AlGaN materials with extremely low thermal activated hole density.

Low resistance (0.1 ~ 1 mΩ cm$^2$) GaN tunnel junctions have been integrated in GaN-based visible light emitters, including visible LEDs[9,12,17-19] and vertical-cavity surface-emitting lasers[14]. Cascaded visible LEDs with multiple active regions interconnected by tunnel junctions were also demonstrated as a method to circumvent the efficiency droop problem.[8,12,19,20] However, achieving low resistance tunnel junctions are challenging for wider bandgap AlGaN materials since interband tunneling probability drops exponentially with increasing barrier height as determined by material bandgap. Moreover, dopant activation energies increase with increasing Al composition, and AlGaN materials typically suffer from a high density of defects and background impurities, which can compensate dopants, making it hard to achieve effective doping, especially for p-type materials.[21] This limits the realization of AlGaN homojunction tunnel junctions. Recent demonstrations using polarization engineering exhibited low tunneling resistances (0.5 ~ 1 mΩ cm$^2$) for Al$_{0.3}$Ga$_{0.7}$N/ In$_{0.25}$Ga$_{0.75}$N[10,16] and Al$_{0.55}$Ga$_{0.45}$N/ In$_{0.2}$Ga$_{0.8}$N[22] tunnel junctions. This indicates the feasibility of tunneling hole injection into UV LEDs using polarization engineered tunnel junctions. In this work, we show that tunnel-injected UV LEDs enable the analysis of compensating impurity concentration in the p-AlGaN layer using capacitance-voltage measurement. We demonstrate efficient tunnel-injected UV LEDs emitting at 325 nm based on optimization of the p-AlGaN cladding layer. Utilizing interband tunneling hole injection, an acceptor free UV LED emitting at similar wavelength was further obtained.

The epitaxial stack of the tunneling injected UV LED structure investigated here is shown in Fig. 1(a). A tunnel junction is formed on top of a conventional UV LED structure, enabling n-type metal contacts to both top and bottom contact layers. The equilibrium fixed and depletion charge profile and band diagrams of the tunneling injected UV LED structures with varying effective dopant density ($N_A^* = N_A - N_{imp}$, where $N_A$ is the acceptor doping concentration, $N_{imp}$ is background compensation charge density) in the p-AlGaN layer are shown in Fig. 1(b). Due to the strong polarization effect, band alignment of the tunnel junction can be achieved by inserting a thin InGaN layer.[9,10,19,22-24] However, the tunneling probability can be greatly reduced by the depletion barrier in the p-AlGaN layer due to the difficulty to achieving degenerate p-type doping, and the large valence band offset between InGaN and AlGaN as shown in the energy band diagram in Fig. 1(b). Increasing the Mg concentration is therefore important to achieve high ionized acceptor (or negative) charge density, and low depletion barrier in the p-AlGaN layer.

To further increase the negative charge density, three dimensional (3D) polarization charges from polarization grading in the p-AlGaN layer can be used, as shown previously by other investigators.[25-28]

Linear grading of Al mole fraction from 75% to 30% in 20 nm p-AlGaN results in ~ $1.5 \times 10^{19}$ cm$^{-3}$ 3D polarization charge, as determined by the gradient of the polarization $\rho = -\nabla \cdot P$. The 3D polarization charge in p-AlGaN allows the formation of a p-n diode even without Mg doping. Field ionization due to polarization grading in p-AlGaN further assists the activation of Mg dopants, leading to near degenerate p-type doping, which is beneficial for enhanced tunneling probability across the tunnel junction.[25]

One challenge to achieving conductive p-type AlGaN lies in the severe compensation of the acceptors due to crystal imperfections including native defects such as N vacancies, and the incorporation of donor-type impurities (e.g. O and/or C).[21] This degrades both the tunnel junction and UV LED active region performance. The band diagrams with different net dopant concentrations ($N_A^* = N_A - N_{imp}$) are compared in Fig. 1(b). Near degenerate p-type AlGaN can be achieved with $N_A^*$ above $5 \times 10^{18}$ cm$^{-3}$ due to the effect of 3D polarization charge. Even at zero net impurity doping ($N_A^* = 0$), the negative polarization charges at the InGaN/p-AlGaN interface and in the graded p-AlGaN layer lead to the valence band being close to the Fermi level with relatively low depletion barrier for tunneling in the p-AlGaN layer.

If the p-AlGaN layer has an effective n-type doping ($N_A^* < 0$), the AlGaN can still function as a hole injection layer for low effective donor concentrations, as shown in the case when $N_{imp} - N_A = 5 \times 10^{18}$ cm$^{-3}$. However, for higher effective donor concentrations (such as $N_{imp} - N_A = 1 \times 10^{19}$ cm$^{-3}$), the entire p-AlGaN layer is depleted. In both these cases, higher voltage would be needed for tunneling.

To study the influence of the donor-type compensating impurities, four tunnel-injected UV LED structures (Fig. 1(a)) with different p-AlGaN layers were grown under similar growth conditions.[10] The tunnel-injected UV LEDs were grown by plasma assisted molecular beam epitaxy (PA-MBE) on metal polar Al$_{0.3}$Ga$_{0.7}$N templates with threading dislocation density of $2 \times 10^9$ cm$^{-2}$. The UV LEDs consisted of 400 nm n-Al$_{0.3}$Ga$_{0.7}$N bottom contact layer, 50 nm n-Al$_{0.3}$Ga$_{0.7}$N cladding layer with [Si] = $3 \times 10^{18}$ cm$^{-3}$, three periods of 2.5 nm Al$_{0.2}$Ga$_{0.8}$N/ 7.5 nm n-Al$_{0.3}$Ga$_{0.7}$N ([Si] = $3 \times 10^{18}$ cm$^{-3}$) quantum wells/ barriers, 1.5 nm AlN electron blocking layer, graded p-AlGaN layer with Al mole fraction grading from 75% to 30%, thin tunnel junction layer with 4 nm $In_{0.25}Ga_{0.75}N$, and 150 nm n-$Al_{0.3}Ga_{0.7}N$ top contact layer. Three samples with 20 nm p-AlGaN were doped with different Mg doping levels, which are 0 (A), $7\times10^{18}$ (B) and $3\times10^{19}$ (C) $cm^{-3}$, respectively. Another sample (D) with 50 nm p-AlGaN and $3\times10^{19}$ $cm^{-3}$ Mg doping density was grown to study the influence of the graded p-AlGaN layer thickness on the device performance. The Mg doping densities were achieved by increasing the incoming Mg flux during growth, and calibrated using secondary ion mass spectrometry (SIMS) analysis.

The bottom contact layer was exposed by device mesa isolation using inductively coupled plasma reactive ion etching (ICP-RIE) with $BCl_3$/ $Cl_2$ chemistry. This was followed by bottom contact deposition of Ti (20 nm)/ Al (120 nm) /Ni (30 nm) /Au (50 nm) metal stack and subsequent annealing at 850 °C for 30 sec. Al (20 nm)/ Ni (20 nm) /Au (80 nm) was then evaporated for top contact, which covers 25%, 37% and 52% of the mesa areas for $50 \times 50$, $30 \times 30$, and $20 \times 20$ $\mu m^2$ devices, respectively. Both bottom and top contact resistances are below $5\times10^{-6}$ $\Omega$ $cm^2$ for the samples. On-wafer electroluminescence and power measurements were carried out at room temperature under continuous wave operation. The light was collected from the top surface of the devices using a calibrated Ocean Optics USB 2000+ spectrometer.[16]

Capacitance-voltage (CV) measurements were carried out by reverse biasing the top contact, with excitation frequency of 1 MHz and amplitude of 30 mV. The measured CV curves of devices with full top contact metal coverage are plotted in Fig. 2(a). Sample A ([Mg] = 0) and B ([Mg] = $7\times10^{18}$ $cm^{-3}$) show nearly constant capacitances with increasing reverse bias especially at low voltage range (< 2.5 V). In comparison, when the p-AlGaN layer is heavily doped to $3\times10^{19}$ $cm^{-3}$ (C and D), the capacitance drops greatly with increasing reverse bias.

The measured capacitance ($C_{total}$) can be modeled as two series capacitances due to the tunnel junction ($C_{TJ}$) and the p-QWs-n junction ($C_{PN}$) as shown in Fig. 1(b). Because of the strong polarization field in the InGaN layer, the tunnel junction capacitance is much larger than the p-n junction capacitance and is expected to remain nearly constant with applied bias. Therefore, the overall capacitance $C_{total} = $ $C_{TJ}C_{PN}/(C_{TJ} + C_{PN})$ can be approximated as $C_{total} \sim C_{PN}$. The p-n junction capacitance ($C_{PN}$) is determined by the effective doping densities ($N_A^*$ in p-AlGaN and $N_D$ in n-AlGaN) and the reverse bias ($V_{rev}$) across the junction, as expressed by $\frac{1}{C_{PN}^2} = \frac{2}{q\varepsilon} \frac{N_A^* + N_D}{N_A^* N_D} (V_{bi} - V_{rev})$, where $V_{bi}$ is the build-in voltage, $\varepsilon$ is the dielectric constant and $q$ is unit charge. Based on the above equation, effective doping density $N_{eff} = \frac{N_A^* N_D}{N_A^* + N_D}$ can be extracted.

As shown in Fig. 2(b), sample A and B show near constant capacitance and depletion width indicating modulation of high density charges. This corresponds to full depletion of the graded p-AlGaN layers, with 2D electron gases originated at the AlN/ $Al_{0.3}Ga_{0.7}N$ or $Al_{0.3}Ga_{0.7}N$/ $Al_{0.2}Ga_{0.8}N$ heterointerfaces in the active region. As determined from the calculated equilibrium band diagrams (Fig. 1(b)), a net donor-type compensating charge density ($N_{imp} - N_A$) of $5 \times 10^{18}$ cm$^{-3}$ is required to deplete the graded p-AlGaN layer by compensating the negative polarization charge. In Sample B, where the p-AlGaN layer was doped to [Mg] = $7 \times 10^{18}$ cm$^{-3}$, the donor-type compensating impurity density can be estimated to be at least $N_{imp} = 1.2 \times 10^{19}$ cm$^{-3}$. This is similar to previously reported compensating impurity density in p-type GaN layer grown by ammonia MBE.[29] Since Sample A does not have any Mg doping, it is also depleted, and shows similar (constant) CV profiles. The small difference in the absolute value of capacitance and depletion width may be due to variations in the growth.

The compensating charge was overcome by heavy doping in sample C and D with [Mg] = $3 \times 10^{19}$ cm$^{-3}$, leading to a net acceptor doping density ($N_A^*$) of $1.8 \times 10^{19}$ cm$^{-3}$. Therefore, the extracted effective doping density can be approximated as $N_{eff} = N_D$ due to much higher net acceptor density in the p-AlGaN layer than the donor density in n-AlGaN ($N_A^* \gg N_D$). Figure 2(b) shows that the effective doping concentrations in both heavily doped samples (C and D) drop with increasing depletion width and stabilize at ~ $3 \times 10^{18}$ cm$^{-3}$, which matches the doping density in the barriers and the n-$Al_{0.3}Ga_{0.7}N$ cladding layer. This indicates depletion in the n-AlGaN layer due to efficient doping achieved in p-AlGaN.

Current-voltage characteristics of the samples are shown in Fig. 3. When the device is reverse biased, the p-QWs-n junction is reverse biased, which blocks the current. When the devices are forward biased, the tunnel junction is reverse biased, leading to tunneling hole injection into the forward biased active region. The tunnel-injected UV LED devices showed similar reverse leakage current, but the turn-on voltage varied with different p-AlGaN layers. For the samples (A, B and C) with 20 nm graded p-AlGaN, the voltages at 20 A/cm$^2$ were 5.98 V (A), 6.22 V (B) and 5.75 V (C), respectively. The lowest voltage was obtained for sample C due to reduced depletion barrier for tunneling resulted from heavy p-type doping. By contrast, because of the absence of acceptors in sample A, the wide depletion barrier in p-AlGaN prevents effective interband tunneling at low bias, leading to severe electron overflow and soft turn-on of the devices. The samples showed a reduction in the overall resistance with increasing Mg doping level as shown in Fig. 3(b). This is attributed to contributions from both reduced p-AlGaN series resistance and tunneling resistance due to increased doping levels.

When p-AlGaN thickness was increased to 50 nm (sample D), the voltage at 20 A/cm$^2$ increased to 6.56 V. The differential resistance was higher than that for sample C as well. The underlying reason is the lower 3D polarization charge density ($\rho_{3D} = 0.6 \times 10^{19}$ cm$^{-3}$) due to increased graded layer thickness, and correspondingly lower polarization field for field-assisted acceptor ionization. This results in wider tunneling barrier as confirmed from CV measurement, and further leads to higher tunneling resistances and higher voltage to turn on the tunnel junction.

On-wafer electroluminescence (EL) and optical power measurements were carried out at room temperature. As shown in Fig. 4, all samples showed single peak emission at approximately 325 nm. This demonstrates effective interband tunneling hole injection through AlGaN/ InGaN tunnel junction. Sample A and B showed highly non-uniform emission from the devices, which is attributed to conduction through low tunneling barrier paths associated with AlGaN and InGaN compositional fluctuations.[30] In comparison, the heavily doped samples C and D showed uniform emission over the entire device region.

The output power showed abrupt increase with increasing Mg doping level from less than 1 μW in the acceptor-free LED (A) to above 1 mW in the heavily doped samples (C and D). Both the external quantum efficiency (EQE) and wall-plug efficiency values were two orders of magnitude higher in the higher doped samples (C and D) than in the low and undoped samples (A and B). The maximum measured power was 1.38 mW at 12 mA, corresponding to 55 W/cm$^2$ at 480 A/cm$^2$. The highest external quantum efficiency of 3.37% was obtained from sample D, while sample C has the highest peak wall-plug efficiency of 1.62%. The efficiency curves did not show saturation and droop for the samples with thin p-AlGaN layers (A, B, C), while the sample (D) with thicker p-AlGaN layer showed saturation near 200 A/cm$^2$. We attribute this to better electron confinement and lower tunneling leakage through the thin electron blocking layer by using thicker p-AlGaN grading as depicted in the upper inset of Fig. 5(a).

The above efficiencies are likely limited by the low light extraction efficiency due to the absence of surface roughening and device packaging, and the low internal quantum efficiency associated with the high defect density in the active region. Secondary ion mass spectrometry (SIMS) measurement of the MBE-grown AlGaN films with either p-type or n-type doping under similar conditions showed that both the oxygen and carbon concentrations were ~ 3×10$^{17}$ cm$^{-3}$, indicating minimal contribution to the compensation effect. Therefore, the high compensation impurity density could come from dislocation related traps, or native defects such as vacancies.[31,32] Optimizing the AlGaN material quality is expected to greatly lower the defect density in both the active region and the p-AlGaN layers.

In summary, we have discussed the design of the p-AlGaN cladding layer towards efficient tunneling injected UV LEDs. Capacitance-voltage analysis is found to be a powerful tool for probing the doping and polarization properties of AlGaN based heterostructure devices. The donor-type compensating impurity density in p-AlGaN layers is estimated to be above 1.2×10$^{19}$ cm$^{-3}$ using capacitance-voltage measurement. Benefiting from the polarization induced 3D charge, an acceptor-free UV LED was achieved with holes injected from interband tunneling. For the UV LEDs emitting at 325 nm, the maximum on-wafer external quantum efficiency and wall-plug efficiency were 3.37% and 1.62%, respectively. Both values are among the highest efficiencies for the reported UV LEDs with similar emission wavelength, confirming the potential of tunneling injection for UV LEDs. This work demonstrates the potential to achieving highly efficient UV LED using interband tunneling hole injection.

Acknowledgement:

We acknowledge funding from the National Science Foundation (ECCS-1408416). Sandia National Laboratories is a multi-program laboratory managed and operated by Sandia Corporation, a wholly owned subsidiary of Lockheed Martin Corporation, for the U.S. Department of Energy's National Nuclear Security Administration under contract DE-AC04-94AL85000.

Figure captions:

Fig. 1 (a) Epitaxial stack of the tunneling injected UV LED structure. (b) Equilibrium fixed and depletion charge profile (mobile charges are not shown) and band diagrams of the tunneling injected UV LED structures with varying effective dopant density ($N_A^* = N_A - N_{imp}$) in the p-AlGaN layer.

Fig. 2 (a) Measured CV curves and (b) extracted effective depletion charges ($N_{eff}$) of the tunneling injected UV LED devices. Sample A and B show near constant capacitance and depletion width indicating modulation of high density charges. Sample C and D show larger increase in the depletion width with reverse bias.

Fig. 3 (a) I-V characteristics and (b) differential resistances of the tunnel-injected UV LED samples.

Fig. 4 EL spectrums and corresponding microscope images of the tunnel-injected UV LED devices.

Fig. 5 (a) Output power, (b) EQE and (c) WPE of the tunnel-injected UV LED devices. The schematic conduction band profiles under device operation for the samples with different p-AlGaN grading thicknesses are shown in the upper inset of (a). Thicker compositional graded p-AlGaN layer could lead to lower tunneling leakage through the thin electron blocking layer. The powers were measured on-wafer from the top surface of the devices without integrating sphere.

References:

1. M. Kneissl and J. Rass, *III-Nitride Ultraviolet Emitters*. (Springer International Publishing, Switzerland, 2016).
2. J. Rass, T. Kolbe, N. Lobo-Ploch, T. Wernicke, F. Mehnke, C. Kuhn, J. Enslin, M. Guttmann, C. Reich, and A. Mogilatenko, SPIE OPTO, 93631K (2015).
3. H. Hirayama, S. Fujikawa, N. Noguchi, J. Norimatsu, T. Takano, K. Tsubaki, and N. Kamata, physica status solidi (a) **206** (6), 1176 (2009).
4. M. Shatalov, W. Sun, R. Jain, A. Lunev, X. Hu, A. Dobrinsky, Y. Bilenko, J. Yang, G. A. Garrett, and L. E. Rodak, Semiconductor Science and Technology **29** (8), 084007 (2014).
5. M. Jo, N. Maeda, and H. Hirayama, Applied Physics Express **9** (1), 012102 (2015).
6. H.-Y. Ryu, I.-G. Choi, H.-S. Choi, and J.-I. Shim, Applied Physics Express **6** (6), 062101 (2013).
7. J. Li, T. Oder, M. Nakarmi, J. Lin, and H. Jiang, Applied physics letters **80** (7), 1210 (2002).
8. F. Akyol, S. Krishnamoorthy, and S. Rajan, Applied Physics Letters **103** (8), 081107 (2013).
9. S. Krishnamoorthy, F. Akyol, and S. Rajan, Applied Physics Letters **105** (14), 141104 (2014).
10. Y. Zhang, S. Krishnamoorthy, J. M. Johnson, F. Akyol, A. Allerman, M. W. Moseley, A. Armstrong, J. Hwang, and S. Rajan, Applied Physics Letters **106** (14), 141103 (2015).
11. A. G. Sarwar, B. J. May, J. I. Deitz, T. J. Grassman, D. W. McComb, and R. C. Myers, Applied Physics Letters **107** (10), 101103 (2015).
12. S. M. Sadaf, Y.-H. Ra, H. P. T. Nguyen, M. Djavid, and Z. Mi, Nano letters **15** (10), 6696 (2015).
13. M. J. Grundmann and U. K. Mishra, physica status solidi (c) **4** (7), 2830 (2007).
14. J. Leonard, E. Young, B. Yonkee, D. Cohen, T. Margalith, S. DenBaars, J. Speck, and S. Nakamura, Applied Physics Letters **107** (9), 091105 (2015).
15. F. Akyol, S. Krishnamoorthy, Y. Zhang, J. Johnson, J. Hwang, and S. Rajan, Applied Physics Letters **108** (13), 131103 (2016).
16. Y. Zhang, A. Allerman, S. Krishnamoorthy, F. Akyol, M. W. Moseley, A. Armstrong, and S. Rajan, Applied Physics Express **9**, 052102 (2016).
17. E. C. Young, B. P. Yonkee, F. Wu, S. H. Oh, S. P. DenBaars, S. Nakamura, and J. S. Speck, Applied Physics Express **9** (2), 022102 (2016).
18. I. Ozden, E. Makarona, A. Nurmikko, T. Takeuchi, and M. Krames, Applied Physics Letters **79** (16), 2532 (2001).
19. F. Akyol, S. Krishnamoorthy, Y. Zhang, and S. Rajan, Applied Physics Express **8** (8), 082103 (2015).
20. S.-J. Chang, W.-H. Lin, and C.-T. Yu, IEEE Electron Device Letters **36** (4), 366 (2015).
21. C. G. Van de Walle and J. Neugebauer, Journal of Applied Physics **95** (8), 3851 (2004).
22. Y. Zhang, S. Krishnamoorthy, F. Akyol, S. Khandaker, A. Allerman, M. W. Moseley, A. Armstrong, and S. Rajan, Device Research Conference (DRC), 2015 73rd Annual, 69 (2015).
23. S. Krishnamoorthy, F. Akyol, P. S. Park, and S. Rajan, Applied Physics Letters **102** (11), 113503 (2013).
24. M. Kaga, T. Morita, Y. Kuwano, K. Yamashita, K. Yagi, M. Iwaya, T. Takeuchi, S. Kamiyama, and I. Akasaki, Japanese Journal of Applied Physics **52** (8S), 08JH06 (2013).
25. J. Simon, V. Protasenko, C. Lian, H. Xing, and D. Jena, Science **327** (5961), 60 (2010).
26. P. S. Park, S. Krishnamoorthy, S. Bajaj, D. N. Nath, and S. Rajan, IEEE Electron Device Letters **36** (3), 226 (2015).
27. J. Simon, A. K. Wang, H. Xing, S. Rajan, and D. Jena, Applied physics letters **88** (4), 042109 (2006).

28  S. Li, T. Zhang, J. Wu, Y. Yang, Z. Wang, Z. Wu, Z. Chen, and Y. Jiang, Applied Physics Letters **102** (6), 062108 (2013).
29  C. A. Hurni, J. R. Lang, P. G. Burke, and J. S. Speck, Applied Physics Letters **101** (10), 102106 (2012).
30  D. Nath, Z. Yang, C.-Y. Lee, P. Park, Y.-R. Wu, and S. Rajan, Applied Physics Letters **103** (2), 022102 (2013).
31  G. Namkoong, E. Trybus, K. K. Lee, M. Moseley, W. A. Doolittle, and D. C. Look, Applied Physics Letters **93** (17) (2008).
32  R. Lieten, V. Motsnyi, L. Zhang, K. Cheng, M. Leys, S. Degroote, G. Buchowicz, O. Dubon, and G. Borghs, Journal of Physics D: Applied Physics **44** (13), 135406 (2011).

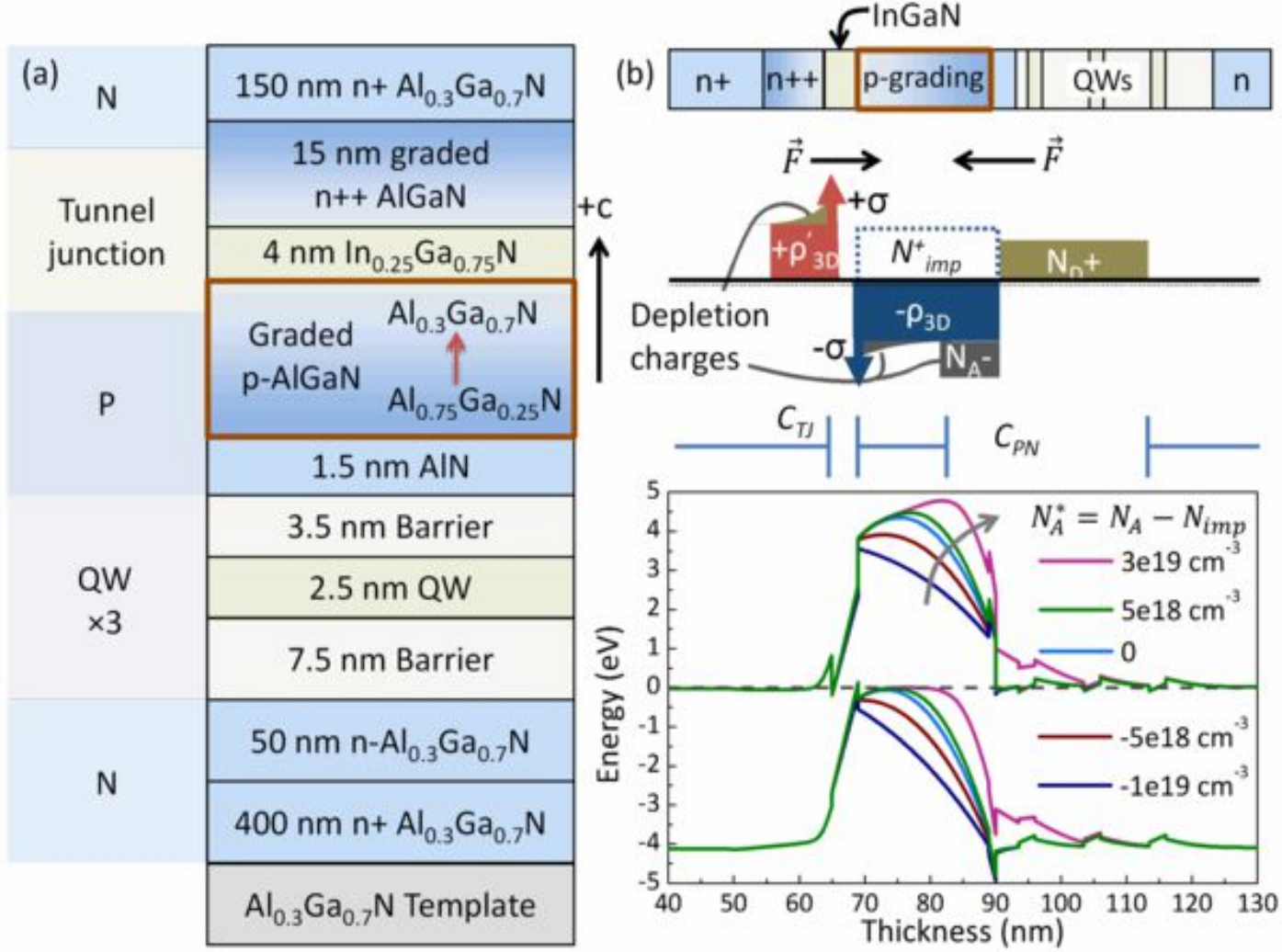

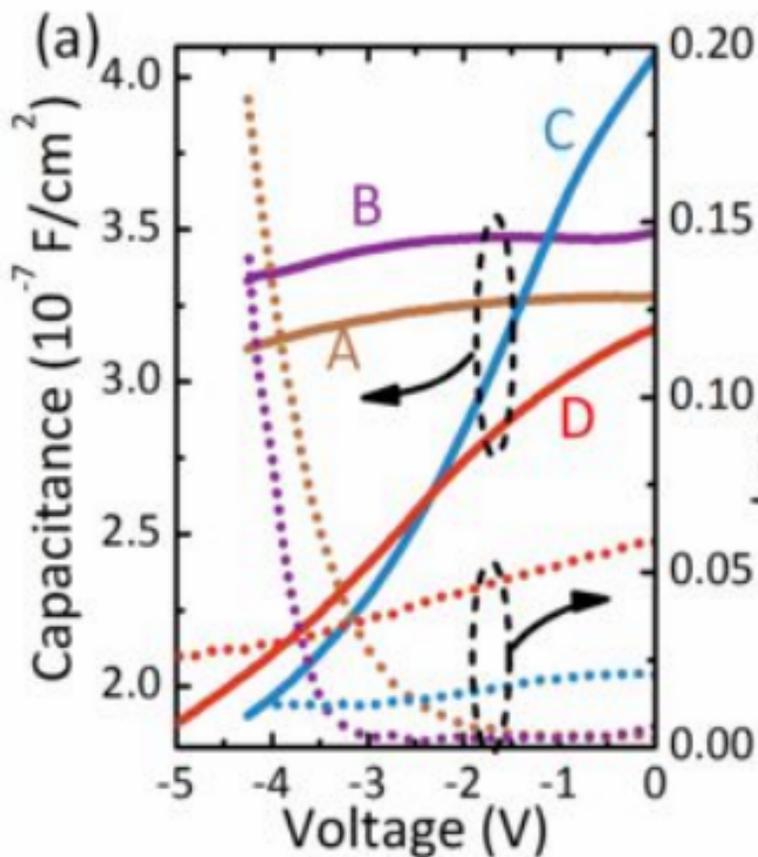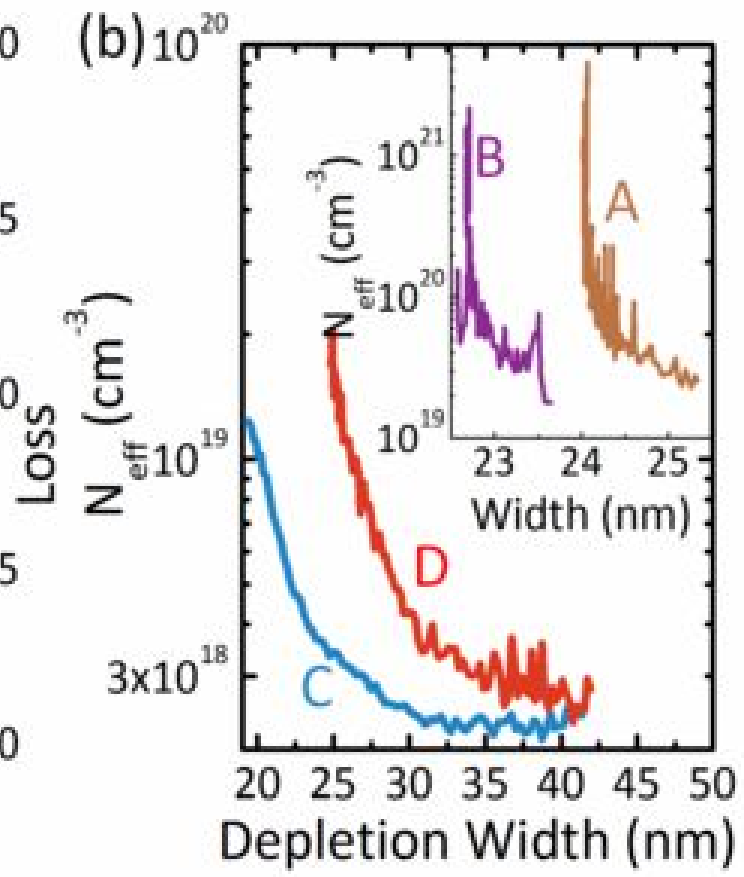

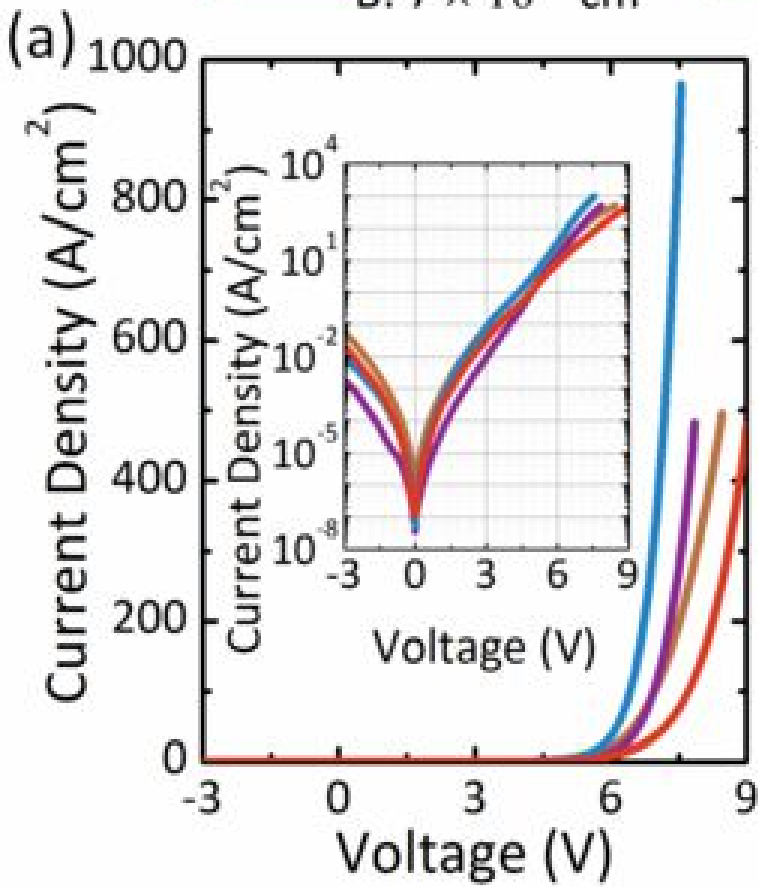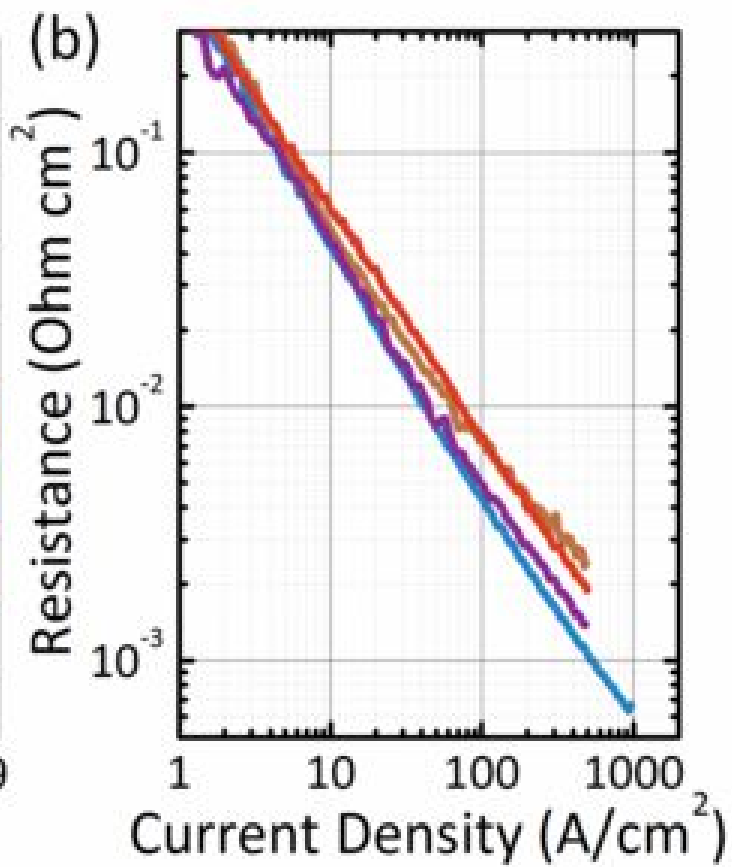

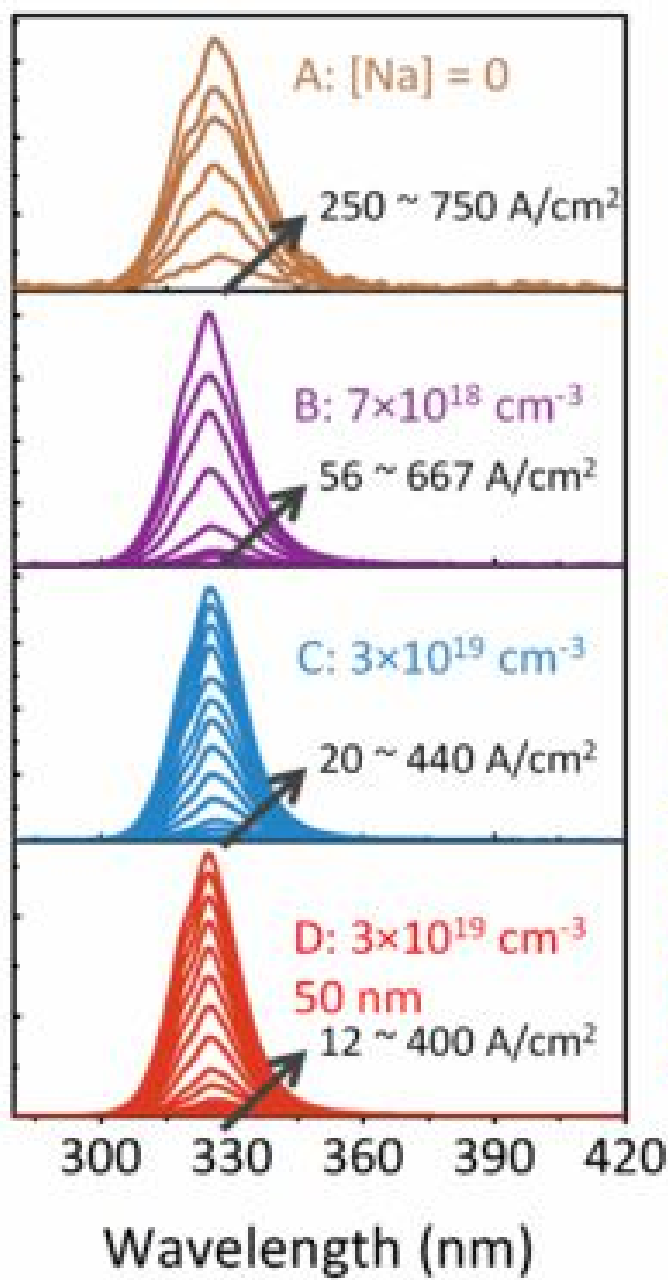

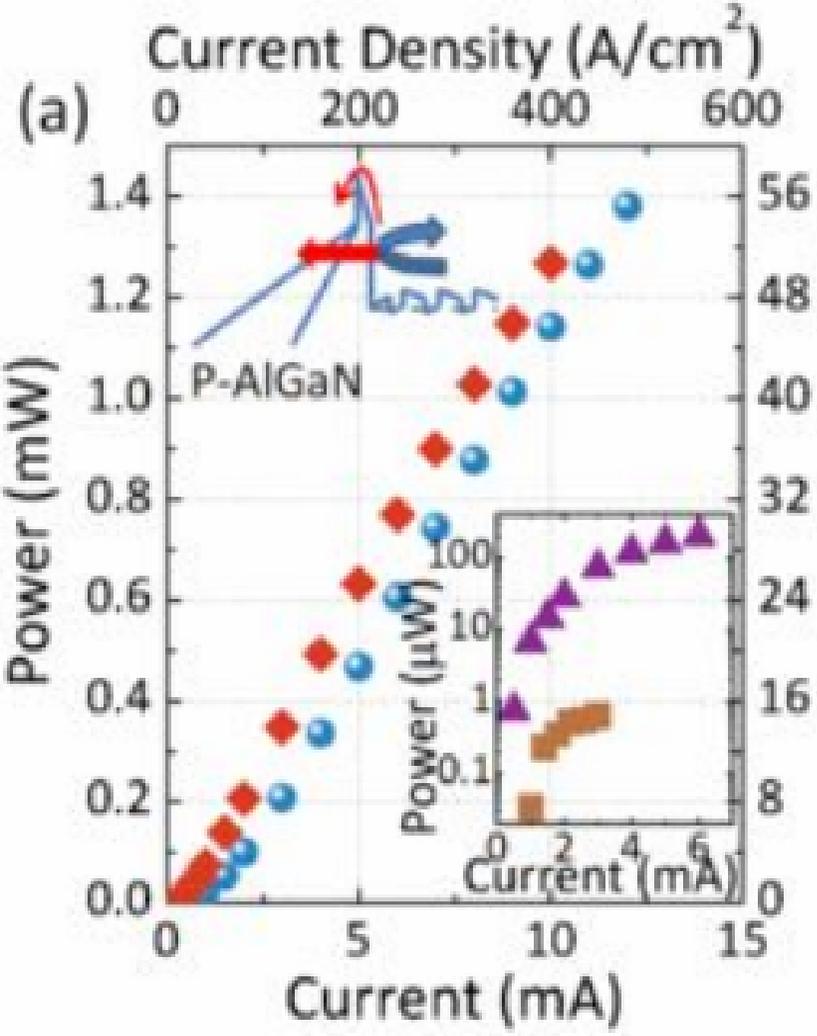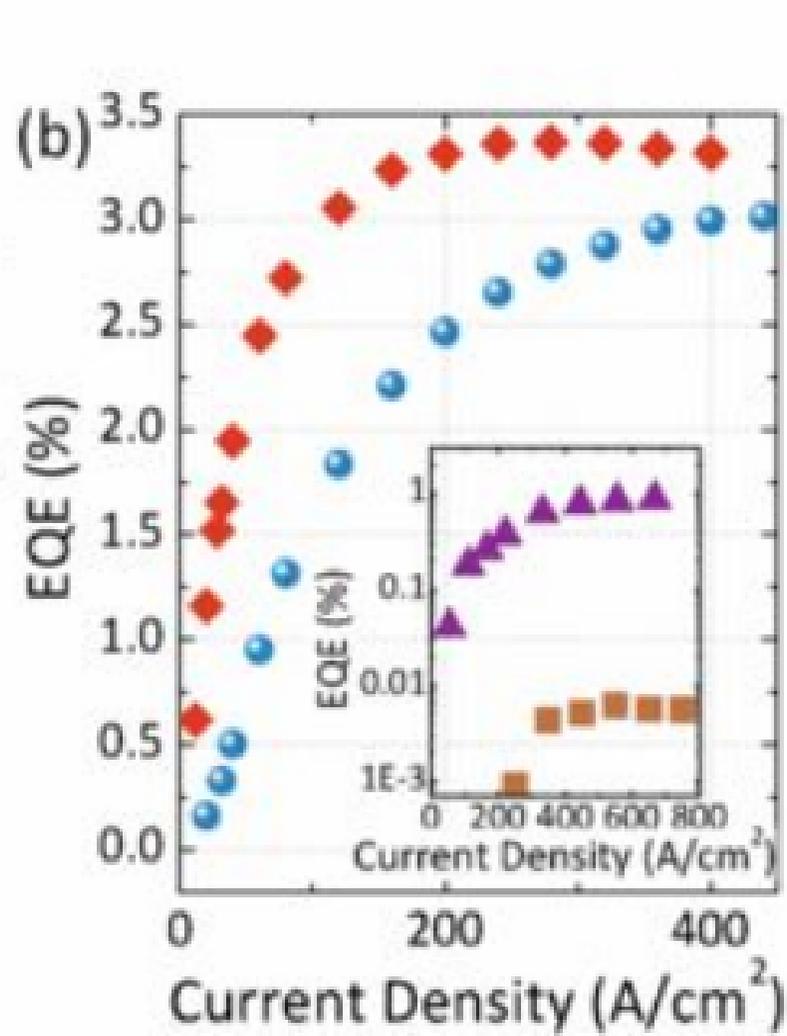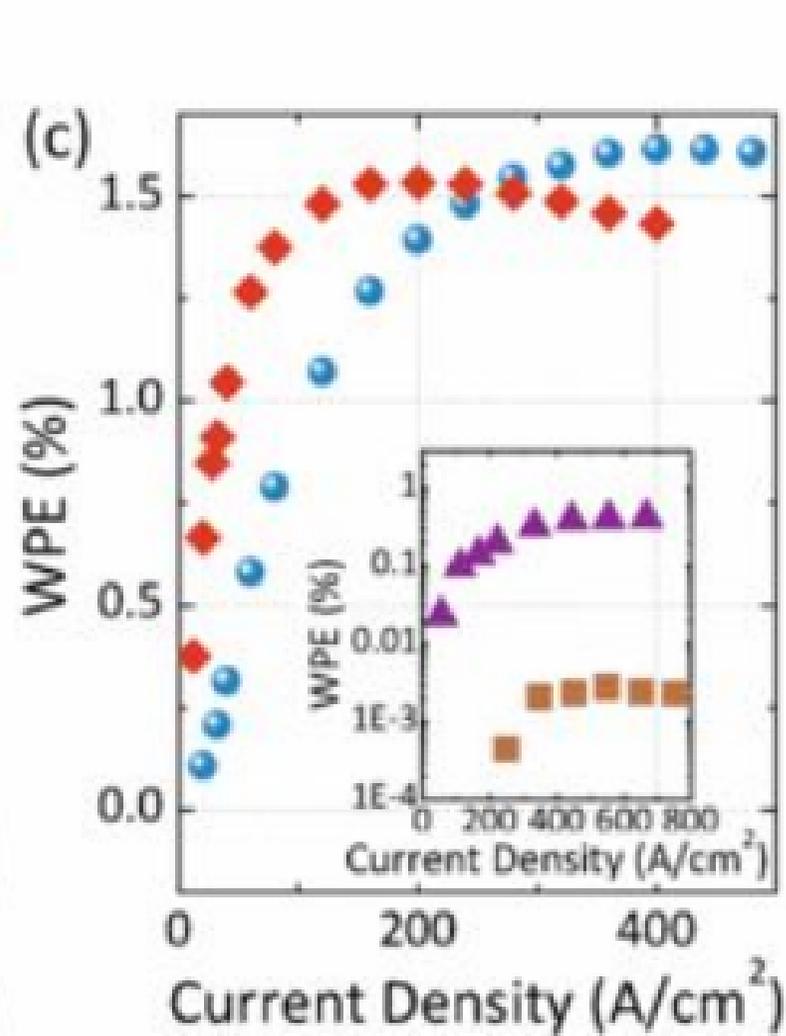

| | |
|---|---|
| ■ | A: Na=0 |
| ▲ | B: $7 \times 10^{18}$ cm$^{-3}$ |
| ● | C: $3 \times 10^{19}$ cm$^{-3}$ |
| ◆ | D: $3 \times 10^{19}$ cm$^{-3}$ 50nm |